# Enhanced Tritium Production in Irradiated TiD$_2$ from Collisional Fusion in the Solid-State

Andrew K. Gillespie [a]*, Cuikun Lin [a], Ian Jones [a], Brad Jeffries [b], Joseph Caleb Philipps [c], Sandeep Puri [a], John Gahl [d], John Brockman [c,e], and R. V. Duncan [a]

AFFILIATIONS

[a] *Department of Physics and Astronomy, Texas Tech University, Lubbock Texas 79409, USA*
[b] *Pacific Northwest National Laboratory, Richland, Washington 99354, USA*
[c] *University of Missouri Research Reactor, Columbia, MO 65211, United States*
[d] *Department of Electrical and Computer Engineering, University of Missouri, Columbia Missouri 65211, USA*
[e] *Department of Chemistry, University of Missouri, Columbia Missouri 65211, USA*
*Author to whom correspondence should be addressed: a.gillespie@ttu.edu

## ABSTRACT

Ongoing research in new nuclear mechanisms hold the potential for beneficial developments in nuclear power cycle designs. Recent reports investigated the possibility of lattice dynamics to influence nuclear processes in metals. Results from Steinetz *et al*., at the NASA Glenn Research Center indicated that it may be feasible to initiate deuterium-deuterium fusion reactions that are enhanced using electron screening to reduce the deuterium-deuterium fusion barrier. This article presents tritium production results from both simulations and experiments targeting specific nuclear processes in an effort to identify the source of higher-energy neutrons observed in those results. We explore two pathways of tritium generation in TiD$_2$ through this fusion cycle. Tritium production from TiD$_2$ in the University of Missouri Research Reactor (MURR), where the neutron spectrum was ~ 90% thermal, was within 25% of the predicted amount from simulations, and well-explained by known nuclear reactions without invoking screening-enhanced recoil-induced fusion. Tritium production from TiD$_2$ in the cyclotron vault at MURR, where the neutron spectrum was completely energetic with almost no thermal neutrons, was a factor of 2.9 – 5.1 times higher than predicted from simulations using known nuclear reactions. This indicates the likelihood of an additional mechanism, such as collision-induced fusion in the solid state, increasing the credibility in the results from Steinetz *et al*.

# I. INTRODUCTION

For decades, researchers and companies have been exploring various concepts for fusion energy, with recent advancements that include the development of high-temperature superconducting magnets, potentially leading to more compact fusion devices.[1] Additionally, in 2022, an experiment at the National Ignition Facility achieved a significant milestone by generating more energy from a fusion reaction than the energy initially input to the nuclear fuel that triggered the reaction.[2-4] In a recent publication, Chen *et al*. observed increased fusion rates in electrochemically loaded metal deuterides.[5] Numerous research groups have also highlighted plasma-electron screening as a crucial physical process to investigate in high-energy-density (HED) experiments. For instance, the National Research Council has noted a heated debate surrounding plasma screening models,[6-9] suggesting that HED experiments could offer valuable insights.

Recently, a novel type of collision-induced, solid-state fusion reaction was reported in titanium deuteride and erbium deuteride by Steinetz *et al*.[10], and theoretically analyzed by Pines *et al*.[11] Their results show that exposing the densely deuterated materials to bremsstrahlung photons resulted in observed energies consistent with the D(d, n)$^3$He reaction. Pines *et al*. proposed that the energetic photo-neutrons from deuteron disintegrations induced from the intense electromagnetic radiation well above the threshold photon energy of 2.23 MeV[10] within the targets (TiD$_2$, ErD$_3$) accelerated some deuterons through photo-neutron collisions. Subsequently, this intense electromagnetic field incident on the samples enhances the screening potential ($U_e$) through induced plasma screening and the Compton-effect electron screening. While conventional fusion typically involves d-beam energy ($E$) significantly surpassing electron screening $U_e$, intriguing possibilities arise when $E$ approaches $U_e$. In such cases, screening effects become pronounced in metals, facilitating d–D fusion ('d' is the energetic deuteron that is accelerated through high-energy neutron collisions in the solid TiD$_2$ and 'D' is the lab-stationary deuteron that the d subsequently hits in the solid), and potentially leading to Oppenheimer-Phillips (OP) reactions with the host lattice ions producing even higher energy neutrons than fusion neutrons. It is worthy to mention that measurement of screening potentials through accelerated d–D collisions in various materials has revealed values around 25 eV in gaseous D$_2$,[12] approximately 50 eV in deuterated insulators, semiconductors, and noble metals (Cu, Ag, Au), roughly 180 eV in Be, and about 800 eV in Pd. These larger screening energy measurements significantly exceed the metal's Fermi energy. Independent groups, including Raiola, Strieder, Czerski, Kasagi, and the Google 'Charleston' collaboration, have corroborated these findings.[13-20,21] Pines *et al*. further analyzed the Steinetz *et al*. experiment, predicting that the elastic neutron collision-accelerated deuteron in D(n,n)d would, on average, possess 64 keV in kinetic energy, resulting in a fusion cross-section of only 17 mb with a lab-stationary deuteron elsewhere in the sample.

According to a careful analysis of their data, Steinetz *et al*. conclude that their e-beam linear accelerator produces the following process in their samples: (1) Gammas initially photo-disassociate D to produce photo-neutrons with an average kinetic energy of 145 keV. (2) These photo-neutrons then scatter elastically with D, with σ = 3 b, to produce d at 64 keV. (3) These d then collide with D to produce the two branches of d–D fusion, with σ = 17 mb.[10,11,22] (4) The resulting 2.45 MeV fusion neutrons collide elastically to produce even higher-energy d, with σ = 2.3 b. (5) These higher-energy d collide with D to produce secondary fusion with σ = 0.1 b. This also may cause some d to accelerate toward erbium or titanium to produce Oppenheimer – Phillips (OP) reactions. It's important to note that the key factor in this reaction is the energetic deuteron resulting from neutron recoil, which can then further react with a lab-stationary D. Therefore, it would be important to understand this new fusion mechanism confined within a lattice, which emphasizes the necessity for thorough examination and replication in experiments. Instead of relying solely on neutrons generated from gammas and X-rays above 2.23 MeV through D photodissociation, it is enlightening to use other available neutron sources for this purpose. The other neutron sources will be a focus of this effort.

There also exist various methods to measure the nuclear byproducts. In typical fusion reactors, the deuterium cycle encompasses two main fusion reactions involving deuterium, each occurring with roughly the same 50% probability:

$$D + D \rightarrow {}^3He + n + 3.27 \text{ MeV} \quad (1)$$
$$D + D \rightarrow T + p + 4.03 \text{ MeV} \quad (2)$$

The $^3$He producing d–D fusion reaction branch (1, above) in plasma generates neutrons with an energy of ~ 2.5 MeV, which serve as a valuable diagnostic tool. For instance, in ohmically-heated tokamaks like Alcator C, where the plasma ion velocity distribution is predominantly Maxwellian, Lehner and Pohl[23] demonstrated that the neutron energy spectrum resulting from a three-dimensional Maxwellian ion distribution follows a Gaussian shape. Several research teams have reported the detection of 2.45 MeV neutrons from various experimental setups.[24,10] The second major d–D reaction results in the production of tritium. As shown in **Figure 1**, the cross-section for the D(d,p)T reaction is approximately 0.077 barns at 1 MeV and 0.017 barns at 0.1 MeV.

In our samples, it is noteworthy that the produced tritium primarily retained as TiDT, or more rarely TiT$_2$, within the material matrix. This unique feature can assist us in designing experiments that significantly improve the signal-to-noise ratio in measurements. Hence, it is of interest to investigate the tritium generation in irradiated samples that have been subjected to varying radiation dose and varying neutron energy spectra to further test these hypotheses.

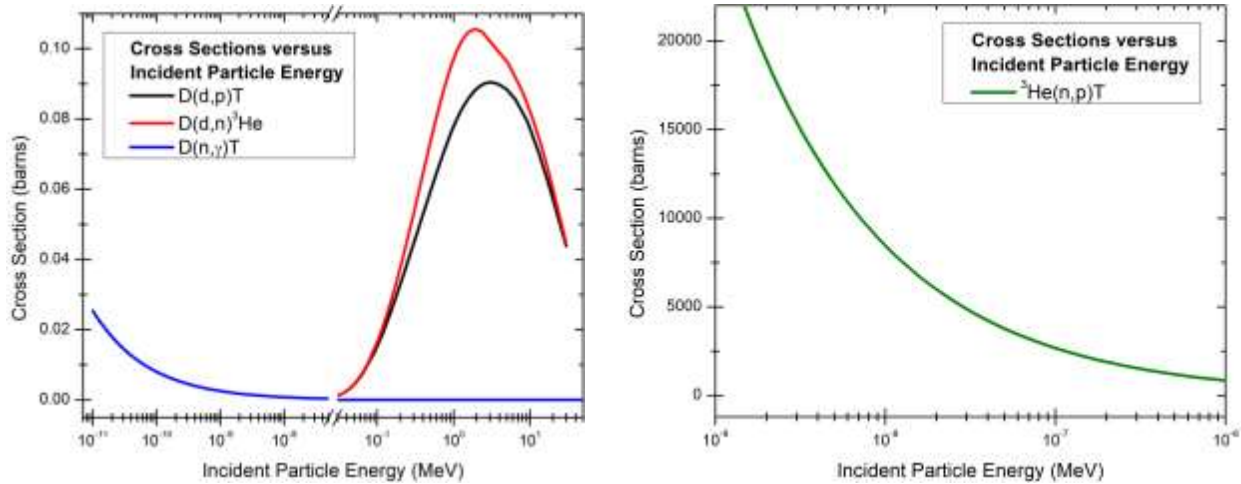

**Figure 1:** Cross sectional data relevant reactions involving incident deuterons (Left) and the thermal energy range of incident $^3$He (Right).[25]

In this paper, we evaluate aspects of this novel approach involving particle collisions within the solid-state lattice to induce nuclear reactions confined to the lattice itself. We test methods of producing energetic tritium in the proposed cycle, or alternatively the collisional acceleration of deuterium by an energetic neutron to adequate energies to initiate collisional fusion reactions with stationary deuterium atoms elsewhere in the solid-state fuel. It is important to note that experiments involving thermal neutrons will introduce alternative known pathways for tritium production. Any $^3$He producing reactions to result in a subsequent $^3$He(n,p)T reaction. Though it has a much lower cross section, radiative capture through D(n,γ)T may also produce tritium.

Additionally, we present Monte Carlo n-Particle (MCNP®) transport simulations[26] and experimental evidence to support our findings. In February, 2022, staff from the NASA Glenn Research Center published an article in IEEE Spectrum[27] that speculated that the emergence of these types of lattice-confined fusion reactions may eliminate the need for the expensive and massive equipment that is currently required to support magnetic or inertial confinement fusion. Further development of this process may eventually result in a practical, solid-state approach to fusion power, which is essential to its ultimate wide-scale utilization.

## II. MATERIAL SYNTHESIS

Bulk chunks of titanium metal were used to make TiD$_2$ and TiH$_2$ powders. The titanium comes from various lot numbers and shipments, each with over 99.9% purity. The loading gases used were research-grade and over 99.999% purity. The titanium was scrubbed in water, dried, and then scrubbed with ethanol. It was placed in a stainless-steel tube under vacuum for about 20 minutes before purging with the loading gas at room temperature. The system was placed under vacuum before the heating profile was started. The heating profile started at room temperature and increased to 750 °C over the course of 1.5 hours. The system was purged with the loading gas once it first reached 100 °C, allowed to dwell for about 5 min, and then placed under vacuum again.

This purging step helped to remove oxygen from the surface. The system was allowed to dwell at 750 °C for one hour before pressurizing the system with 150 torr of pure $D_2$ gas. It was allowed to dwell at this temperature and pressure for 12 hours and then was allowed to return to room temperature. Two large batches were used for this experiment.

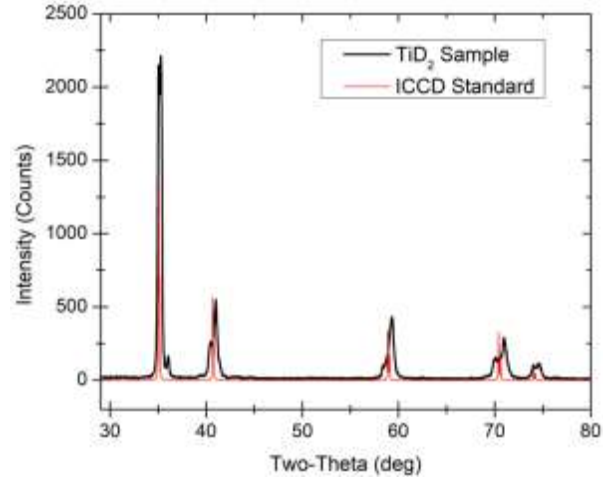

**Figure 2:** The XRD pattern of the synthesized $TiD_2$ sample.

**Figure 2** shows the x-ray diffraction (XRD) results for a titanium deuteride sample. XRD measurements were performed on each sample to confirm the level of purity. This spectrum closely matches the standard data card 04-016-3846 from the International Centre for Diffraction Data,[28,29] thus confirming the successful synthesis.

### III. NEUTRON SOURCES

MURR provides several well-characterized neutron irradiation positions for experimentation. The position used for our first set of irradiation experiments offers a high neutron flux region within the reactor, approximately $(8.21 \pm 0.12) \times 10^{13}$ n/cm²/s, with 90% of neutrons being thermal (E ~ 26 meV) and 10% fast (E > 100 keV). The neutron spectrum, illustrated in **Figure 3**, is typical for a fission neutron spectrum in light water reactors.[30,31] The fast neutron spectrum follows a Watt distribution with function parameters set to a = 1, b = 2, and c = 0.485.

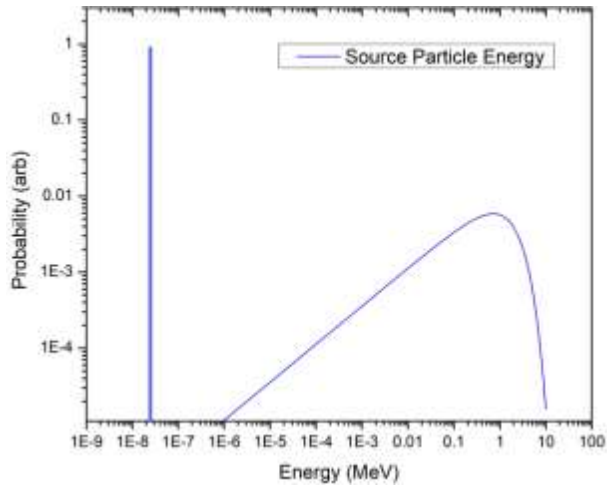

**Figure 3:** Spectrum of neutrons at the MURR reactor port. *Left*: The neutron energy spectrum on a log scale to highlight that 90% of the source consists of thermal neutrons. 10% of the total flux is due to a Watt distribution of fast neutrons.

In the first set of experiments, we have attempted to confirm this reported collision-induced, lattice-confined fusion mechanism using neutron irradiations of $TiD_2$ (and $TiH_2$ controls) within the MURR facilities, rather than the accelerator-driven photo-neutron irradiation used in Steinetz, *et al*. The samples were irradiated for 50 hours over their approximate area of $3.14 \times 10^{-2}$ cm$^2$ for a total dose of $4.64 \times 10^{17}$ neutrons. and quantitative tritium measurements following extraction from the samples after irradiation were used to measure the rate of any neutron collision-induced d–D fusion in the $TiD_2$ samples.

The gamma radiation level in MURR is smaller than in Steinetz, *et al*., but reactor gamma flux / spectrum is uncertain. The gamma radiation level and spectrum has not been measured in MURR for this experiment, and it is highly variable during the reactor cycle. Estimates predict that the gamma flux is between $10^9$ and $10^{11}$ γ/cm$^2$/s when scaling the expected gamma flux from other measured reactors.[32-35] This is below 1% of the measured gamma flux in Steinetz *et al*. The authors note that the high gamma flux could be a significant contributing factor in the NASA-Glenn experiments, These screening effects have been predicted at very low energies, and present interesting physics, but will not enhance d–D fusion screening at energies above ~20 keV significantly.[22] Consequently, the gamma flux in our neutron sources will not be a significant source of increased screening effects within this study.

Another neutron source is the MURR cyclotron facility, which produces energetic neutrons at a rate of $3.1 \times 10^9$ n/cm$^2$/s within the radiation vault during the production of various medical radioisotopes. This method of neutron production is preferred over MURR reactor irradiations because it avoids the production of thermal or epi-thermal neutrons, as confirmed by Jeffries *et al*.[36] An example of the neutron energy spectrum from this facility is depicted in **Figure 4**.

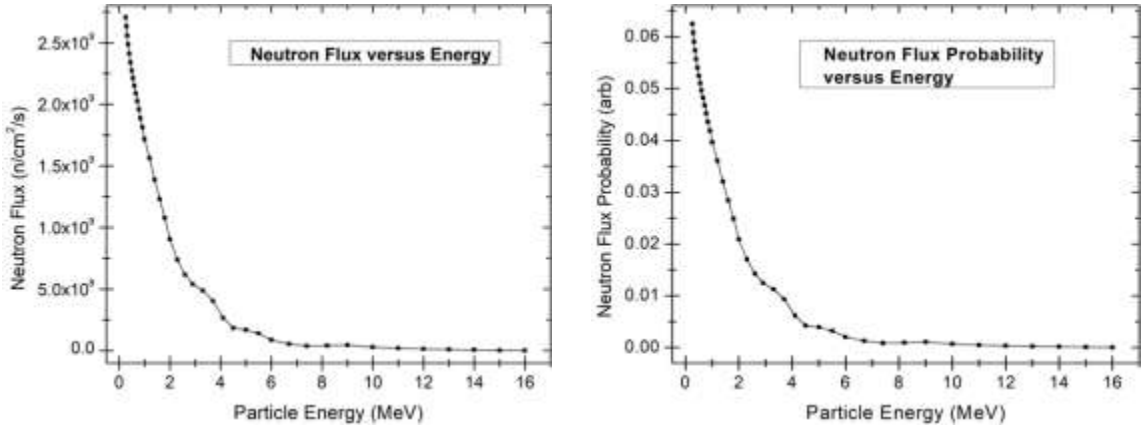

**Figure 4:** Spectrum of neutrons at the MURR cyclotron vault. *Left*: The neutron flux versus energy for the MURR cyclotron vault. *Right*: The normalized neutron flux as a probability used in defining the source in the MCNP® simulations.

This lack of a thermal neutron flux greatly reduces the uncertainty of the level of collision-induced fusion that occurred in the MURR reactor irradiations. We have irradiated both $TiD_2$ experimental samples and $TiH_2$ control samples for at least one month of total irradiation time within this cyclotron vault. The irradiation time for each sample is documented in **Table 1** and **Table 2**. The second set of samples was intermittently irradiated in the cyclotron vault for 260 hours from October 17th, 2022, through December 21st, 2022, for a total dose of $2.04 \times 10^{16}$ neutrons. The third set of samples was irradiated for 376 hours from October 17th, 2022, through January 20, 2023, for a total dose of $2.95 \times 10^{16}$ neutrons.

## IV. MONTE CARLO SIMULATIONS

Each of the experiment types were simulated using Monte Carlo n-Particle calculations. The MCNP® geometries and source conditions were designed to closely match the experimental parameters. For the first set of simulations and experiments, the neutron spectrum from the MURR reactor was used as the source. The source neutrons were incident upon a quartz ampule filled with the metal hydride samples. These ampules were modeled as simple cylindrical shells with a height of 3.2 cm, an outer radius of 0.3 cm, and a thickness near 0.1 cm. The model involved a much smaller $TiD_2$ sample volume of 0.027 cm$^3$ and mass of near 0.10 g within the quartz ampule. The source was modeled as a surface source with an energy probability assigned to each particle history according to the spectrum shown in **Figure 3**. The number of neutrons entering the sample zone was tabulated along with the number of tritium nuclei created. For simulations involving $TiD_2$ samples, about 8.5 tritium nuclei were generated per million source neutrons entering the sample region. When scaling this tritium generation by the total 50-hour irradiation time, it predicted that $(3.9 \pm 0.9) \times 10^{12}$ tritium nuclei would be generated during the experiment.

When simulating the second and third sets of experiments, the source neutrons were incident upon aluminum tubes with cold-welded ends that were filled with the metal deuteride samples. These pinched aluminum tubes were modeled as cylindrical shells for the main shaft with wedged ends that were bound by the outer-most cylinder of the main shaft. The geometry of the tubes and samples is displayed in **Figure 5**.

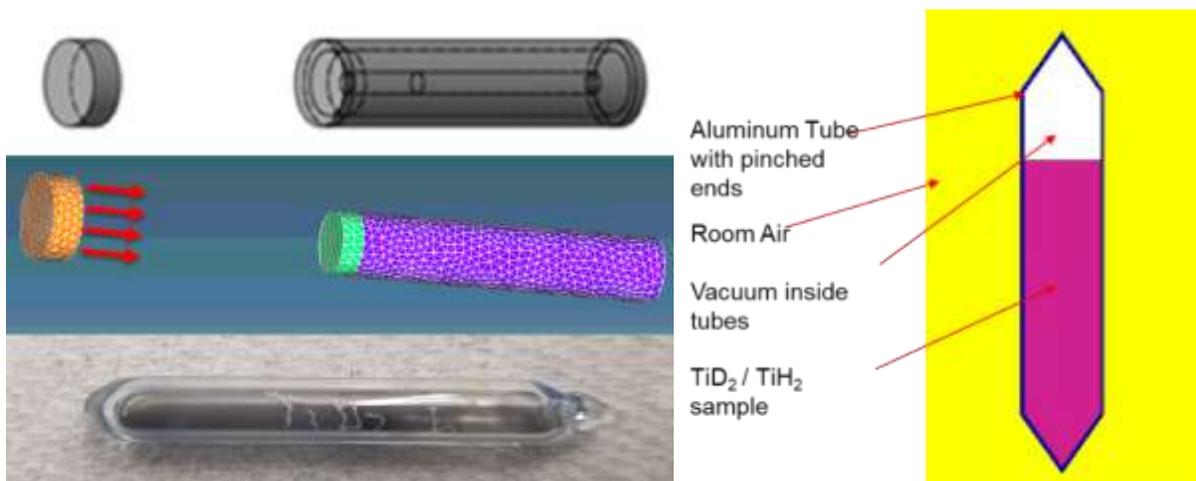

**Figure 5:** *Left*: The simplified simulation geometry and experimental quartz tubes filled with $TiD_2$ used in the first set of experiments within MURR. *Right*: A depiction of the pinched aluminum tube containing a $TiD_2$ sample with vacuum in the headspace. This geometry was used for the second and third sets of experiments and simulations involving the MURR cyclotron vault.

The simulation geometry involves an aluminum tube with pinched ends. The height of the tube is around 7 cm with a radius of 0.64 cm. The $TiD_2$ sample occupies a volume of 5.77 cm$^3$ from the bottom of the tube up to a height of 5 cm. The sample mass is around 15.1 g with a packing fraction of 70%. A rectangular surface source was used with a slightly larger size than the dimensions of the outside of the aluminum tube.

The source was modeled as a surface source with an energy probability assigned to each particle history according to the spectrum shown in **Figure 4**. The number of neutrons entering the sample zone was tabulated along with the number of tritium nuclei created. For simulations involving $TiD_2$ samples, about $0.501 \pm 0.007$ tritium nuclei were generated per million source neutrons entering the sample region. When scaling this tritium generation by the total irradiation time, simulations predicted that about $(1.023 \pm 0.014) \times 10^{10}$ tritium nuclei would be generated for the second set of experiments and about $(1.480 \pm 0.020) \times 10^{10}$ tritium nuclei would be generated during the third set of experiments. The uncertainties reported for the simulations are based on a tritium creation when using a packing fraction range of 69% – 71%.

## V. GAS EXTRACTION AND TRITIUM MEASUREMENTS

All $TiD_2$ and $TiH_2$ samples were synthesized, measured, and analyzed at Texas Tech University (TTU). Tritium concentration measurements were performed as the primary diagnostic since the amount of tritium produced in our metallic samples indicates the fusion efficiency. Tritium extraction and radiation metrology have been developed for quantitative measurements of extraction efficiency.[37,38] Simulations using MCNP® were performed and compared with experimental yields. Tritium concentrations were measured with the Perkin Elmer liquid scintillation analyzer, Quantulus™ GCT 6220,[39,40] which has an uncertainty of only $10^{-14}$ mole. Tritium was recovered in water and dissolved in scintillation fluid for quantitative analysis. When tritium is produced within a metal hydride in this work, it is placed in a hermetic gas manifold and heated to its decomposition temperature. The resulting hydrogen gas is concentrated and reacted with oxygen to form water. The tritium-rich water is collected via washing at least five times with purified water. The wash water is mixed with scintillation liquid and analyzed in the Quantulus. Tritium extraction efficiency is measured to within ±1% accuracy at each step. Typical efficiencies for gas collection, water formation, and tritium detection are 60%, 90%, and 60%, respectively, with an overall recovery efficiency of 33±4%. The efficiency of tritium-containing water collection was determined by decomposing the irradiated metal deuterides and analyzing the wash water with infrared spectroscopy. The wash efficiency is defined as the ratio of the measured $D_2O$ concentration to maximum possible $D_2O$ from the initial known amount of deuterium in the thermally decomposed metal deuteride. This ratio is the tritium recovery efficiency, which assumes that the tritium recovery from the solid matrix is the same as that of deuterium.[41] The tritium detection efficiency is determined during the calibration. A control sample of HPLC water was always measured along with the experimental samples. No tritium was detected in the control group.

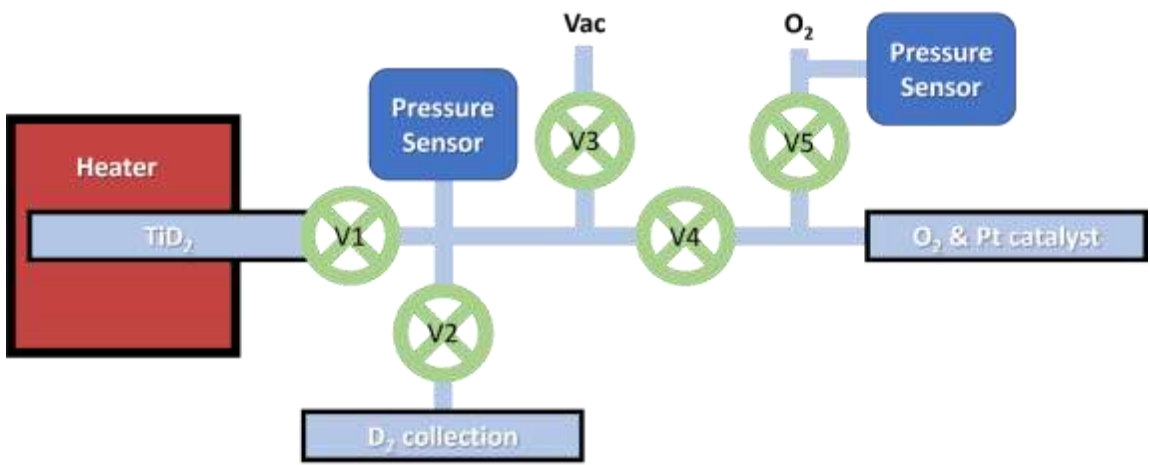

**Figure 6:** A diagram of the gas manifold used in the tritium extraction procedure.

The tritium collection process involves five known volumes. Temperatures and pressures are monitored within all zones during the experiment. First, a hard vacuum is applied to evacuate the entire manifold. With only valve #1 open, as shown in **Figure 6**, the sample chamber is heated past the decomposition temperature to liberate any deuterium, protium, tritium, and any other residual gasses. With valves #1 and #2 open, the $D_2$ gas collection chamber is cooled using liquid nitrogen to capture as much of the non-reacted gas as possible. Valve #1 is closed before allowing the sample chamber to return to ambient temperature. The oxygen chamber is filled such that the oxygen has a stoichiometric ratio of 1:1 with the unreacted gas, assuming that nearly 100% of the gas is either protium or deuterium. This ratio ensures a quick and complete reaction. With valves

#1 and #5 closed, valve #4 is opened. The gas is allowed to diffuse and react on the platinum catalyst to complete the isotope exchange. A known amount of water is used to rinse the chamber to collect the maximum amount of sample fluid as possible. This rinsing step is performed multiple times, and each aliquot is measured separately for tritium activity until no signal is observed.

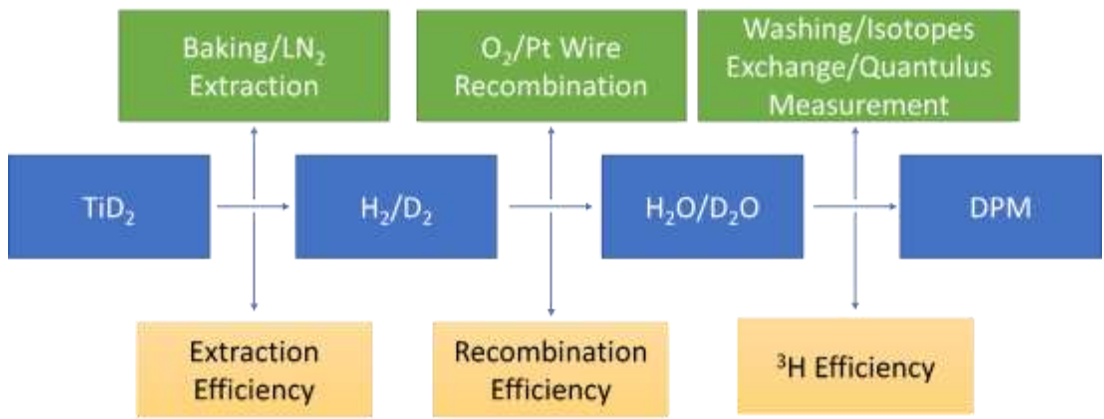

**Figure 7:** Block diagram of the gas extraction method and measurement steps.

There are three steps that must be considered when calculating the extraction efficiency. These steps are outlined in the block diagram in **Figure 7**. During each transfer step, a calculable fraction of the total available gas is collected. When the sample is heated, a small fraction of the total gas is left within the hot sample chamber. When the sample gas is reacting on the catalyst, a fraction of the total gas is left without recombining. During the isotope exchange and tritium measurements, there is an unavoidable, but calibrated quantum efficiency for the detection of tritium.

The scintillator is operated to count the nuclide activity within each sample. This measurement method relies on the interactions between the radioactive materials and a scintillator. The intensity of light is proportional to the initial energy of the beta particle. For every sample assay within the instrument, it measures the activity in a sample vial in units of counts per minute, determines the quench level through the quench-indicating parameters, determines the counting efficiency from the quench curve as a calibration, and calculates the dose per minute for the sample. This dose per minute is equal to the measured counts per minute divided by the quantum efficiency. The calibration standard of Ultima Gold scintillation cocktail 6013681[42] is measured before and after each sample measurement to ensure instrument stability. **Equations 3-5** display the method of converting the measured activity to the number of tritium nuclei. This calculation is performed for 1 Bq, (1 DPS). 1 DPM is equivalent to $9.3 \times 10^6$ nuclei. **Figure 8** displays an example measurement spectrum from the isotope exchange of a $TiD_2$ sample.

$$N = -\frac{dN/dt}{\lambda} = \frac{10^6/sec}{\lambda} \qquad (3)$$

$$\lambda = \frac{0.693}{t_{1/2}} = \frac{0.693}{(12.33yr)(\pi \times 10^7 sec/yr)} = 1.789 \times 10^{-9} s^{-1} \qquad (4)$$

$$N = \frac{1/sec}{1.789 \times 10^{-9}/s} = 5.59 \times 10^8 \, nuclei \qquad (5)$$

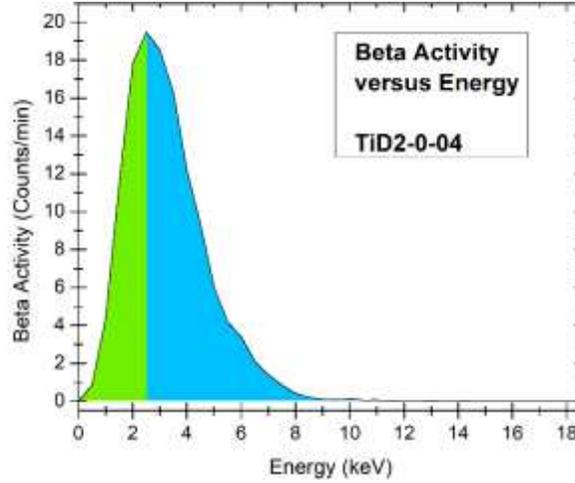

**Figure 8:** The activity spectrum from the first washing step for a titanium deuteride sample that had been irradiated for two minutes. Results for each sample consists of several aliquots from the washing steps.

Lower activity standards, such as the Ultima Gold LLT used in this study, are appropriate for when the scintillator produces low energy events. Lower activity standards require longer collection times to reduce spillover events. No spillover events were observed when collecting these activity spectra, as shown in **Figure 8** with a single activity peak. This is a characteristic beta emission spectrum for tritium. The colors represent counting regions where green is 0-2 keV and cyan is 2-18.6 keV. Experimentally, the region between 0-18.6 keV was counted. This shows that beta decay being measured only originates from tritium and no other decay products.

## VI. RESULTS AND DISCUSSION

### EXPERIMENTS IN THE UNIVERSITY OF MISSOURI RESEARCH REACTOR

Attempts were made to replicate the results of Steinetz, *et al*. to explore whether or not fast n collisions accelerate d to create d–D fusion reactions. In the first set of experiments, samples were irradiated for various durations inside the MURR reactor. Accelerated deuterons can produce D(d,p)T reactions, and we measured tritium quantitatively. **Table 1** shows the measured dose and calculated number of tritium nuclei present in each sample.

**Table 1:** Measured disintegrations per minute and total tritium generation within samples irradiated in the reactor.

| Sample & Irradiation Time | Disintegrations per Minute from Quantulus | Total Tritium Nuclei Produced | Simulated Tritium Nuclei Expected |
|---|---|---|---|
| TiH2-0-01, 2 min | Below detection limit | Below detection limit | 0 |
| TiH2-0-02, 4 min | Below detection limit | Below detection limit | 0 |
| TiH2-0-03, 6 min | Below detection limit | Below detection limit | 0 |
| TiD2-0-04, 2 min | 117 | $3.3 \times 10^9 \pm 3.4 \times 10^7$ | $2.6 \times 10^9 \pm 6.5 \times 10^8$ |
| TiD2-0-05, 4 min | 236 | $7.6 \times 10^9 \pm 7.8 \times 10^7$ | $5.2 \times 10^9 \pm 1.3 \times 10^9$ |
| TiD2-0-06, 6 min | 412 | $1.2 \times 10^{10} \pm 1.2 \times 10^8$ | $7.8 \times 10^9 \pm 1.9 \times 10^9$ |
| TiD2-1-07, 50 h | 186879 | $5.0 \times 10^{12} \pm 5.1 \times 10^7$ | $3.9 \times 10^{12} \pm 9.0 \times 10^{11}$ |
| TiD2-1-08, 50 h | 208963 | $5.8 \times 10^{12} \pm 6.0 \times 10^7$ | $3.9 \times 10^{12} \pm 9.0 \times 10^{11}$ |
| TiD2-1-09, 50 h | 113843 | $5.3 \times 10^{12} \pm 5.4 \times 10^7$ | $3.9 \times 10^{12} \pm 9.0 \times 10^{11}$ |
| TiD2-1-10, 50 h | 146110 | $5.6 \times 10^{12} \pm 5.8 \times 10^7$ | $3.9 \times 10^{12} \pm 9.0 \times 10^{11}$ |

The conclusion from the 50-hour reactor irradiation experiments is that between $5.0 \times 10^{12}$ – $5.8 \times 10^{12}$ tritium nuclei were produced in the original samples based upon the best available characterization of the reactor's neutron spectrum and flux. Simulations predicted $(3.9 \pm 0.9) \times 10^{12}$ tritium nuclei were expected to be produced, though simulation results can be significantly influenced by the orientation of the sample relative to the source. For this first set of irradiation experiments, a 25% uncertainty was assigned to simulated estimates due to an unknown orientation of the sample compared to the fast neutron flux. The measured amounts of tritium had approximately a 25% difference with the simulated estimates. Any significant deviation between the experimental results and the simulated estimations might have indicated additional induced collisional fusion. However, the level of tritium production that we measured was consistent with the known reactions discussed above. Therefore, we were unable to confirm the existence of additional fast-neutron induced d–D collisional fusion tritium production from these data. Hence, we did not see evidence of Steinetz, *et al.*'s results in this initial experiment. This shows that the amount of tritium produced in these high-neutron-flux MURR irradiation experiments is explained by known physical processes without invoking screening-enhanced neutron-recoil-induced reactions.

Additionally, if any $^3$He was formed through the D(d,n)$^3$He pathway, then the $^3$He would rapidly burn up into tritium through the $^3$He(n,p)T reaction due to the high thermal neutron flux. So, no evidence of $^3$He production was detected following the irradiation, and none was expected. This was confirmed through careful measurements using our advanced and quantitatively calibrated FT-ICR mass spectrometer systems.[43]

*EXPERIMENTS IN THE MURR CYCLOTRON VAULT*

The second and third sets of experiments were performed using the MURR cyclotron as our fast neutron source. Samples were placed at the location of their TS-1700 target using a 72 µA current. **Table 2** shows the dose per minute, and number of tritium nuclei present in each sample. The reported uncertainties are based on the time uncertainty in removing the samples from the irradiation vault and the time of measurement. Two days of uncertainty were used with the half-life of tritium to determine the uncertainty in the tritium amount. The first two samples in **Table 2** were measured within 8 months of the irradiation experiment, so approximately 96% of the original tritium amount remained at the time of the measurement. For the third sample in **Table 2** named "TiD2-3-13," over two years elapsed between the irradiation and the tritium measurement, so approximately 88% of the original tritium amount remained at the time of the measurement. At that time, the activity reduces by approximately 0.014% per day. Therefore, an uncertainty of 0.028% was applied to the measured amounts. An additional uncertainty source of 1% was included due to quantum efficiency and the quench calibration curve.

**Table 2:** Measured disintegrations per minute and total tritium generation within samples irradiated in the cyclotron vault.

| Sample & Irradiation Time | Disintegrations per Minute from Quantulus | Total Tritium Nuclei Produced | Simulated Tritium Nuclei Expected |
|---|---|---|---|
| TiD2-2-11, 260 h | 149.5 | $4.2 \times 10^{10} \pm 4.3 \times 10^{8}$ | $1.023 \times 10^{10} \pm 1.4 \times 10^{8}$ |
| TiD2-2-12, 260 h | 97.8 | $3.0 \times 10^{10} \pm 3.1 \times 10^{8}$ | $1.023 \times 10^{10} \pm 1.4 \times 10^{8}$ |
| TiD2-3-13, 376 h | 391.4 | $7.5 \times 10^{10} \pm 7.7 \times 10^{8}$ | $1.480 \times 10^{10} \pm 2.0 \times 10^{8}$ |

The conclusion from the cyclotron vault irradiation experiments is that about $3.0 \times 10^{10}$ – $4.2 \times 10^{10}$ tritium nuclei were produced in the second set of experimental samples and $7.5 \times 10^{10}$ tritium nuclei were produced in the third set of experimental samples. These results were compared to the MCNP® transport calculations using the same source spectrum as the physical experiment and a simplified geometry for the sample tubes. Those simulations predicted that about $(1.023 \pm 0.014) \times 10^{10}$ tritium nuclei were expected to be generated from the second set of experiments and about $(1.480 \pm 0.020) \times 10^{10}$ tritium nuclei were expected to be produced from the third set of experiments. The experimental results yielded 2.9 – 5.1 times the tritium production than the MCNP® simulations predicted. These results, in which only high-energy neutrons were present, in comparison with the reactor measurements (summarized above) where the neutrons were primarily thermal, suggest that some other mechanism is present in producing tritium with high-energy neutrons. Certainly, the collisional acceleration model from Steinetz, *et al.*, is one possibility for the origin of this additional tritium.

The experiments indicate that the actual cross sections for fusion may be slightly higher than those recorded thus far. Though this study offers insight into the underlying mechanisms that could be responsible for the lattice-confined d–D fusion observed by Steinetz, *et al.*, additional experiments are necessary to replicate our results.

**VII. CONCLUSIONS**

Attempts to replicate the results of Steinetz *et al*. were conducted by irradiating samples in the MURR reactor and cyclotron vault to explore fast neutron-induced d–D fusion reactions. Samples irradiated in the reactor yielded between $5 \times 10^{12}$ and $6 \times 10^{12}$ tritium nuclei, consistent with simulations of known physical processes. Additional fusion events might have occurred, but resulting byproducts were below the detection limit for tritium. The high thermal neutron flux facilitated tritium generation through the D(n,γ)T reaction. Samples irradiated in the cyclotron vault yielded a tritium production rate that was between 2.9 – 5.1 times larger than the predicted rates from simulations. The high gamma flux measured in previous experiments by Steinetz, *et al*. could be a significant contributing factor in observed d–D fusion events, and future experiments will explore the effects of the gamma environment on this rate. Based upon the much higher rate of tritium production when only high-energy neutrons were present, we conclude that some other mechanism may be responsible for tritium production in this case, consistent with the Steinetz, *et al*. hypothesis.

Improvements in our understanding of new nuclear mechanisms has the potential to accelerate new power cycle designs. An understanding of increased fusion reaction cross sections due to the effects of electron screening could enable lower temperature reactors with solid state material replacing exclusively plasma-based systems .

## VIII. ACKNOWLEDGEMENTS

This work was supported by Department of Energy award No. DE-AR0001736, the Texas Research Incentive Program, and by Texas Tech University. The identification of commercial products, contractors, and suppliers within this article are for informational purposes only, and do not imply endorsement by Texas Tech University, their associates, or their collaborators. The authors would like to thank the Texas Tech physics shop for their technical assistance and useful discussions. We are grateful to Matthew Looney (Texas Tech) for his suggestions and his efforts as one of the Radiation Safety Officers involved in this effort.

## IX. DATA AVAILABILITY

The data and methods recorded and utilized in this study can be found at https://www.depts.ttu.edu/phas/cees/,[44] and through the Information Technology Division of Texas Tech University.[45] Other data that support the findings of this study are available from the corresponding author upon reasonable request.

## X. DECLARATION

The authors declare that the research was conducted in the absence of any commercial or financial relationships that could be construed as a potential conflict of interest.

# XI. REFERENCES

1. MIT News, Tests show high-temperature superconducting magnets are ready for fusion. (March 4, 2024). Commonwealth Fusion Systems. Retrieved April 25, 2024. https://news.mit.edu/2024/tests-show-high-temperature-superconducting-magnets-fusion-ready-0304.

2. A shot for the ages: Fusion ignition breakthrough hailed as 'one of the most impressive scientific feats of the 21st century' (December 14, 2022). Lawrence Livermore National Laboratory. Retrieved May 6, 2024, https://www.llnl.gov/article/49301/shot-ages-fusion-ignition-breakthrough-hailed-one-most-impressive-scientific-feats-21st.

3. Nature. US nuclear-fusion lab enters new era: achieving 'ignition' over and over (December 15, 2023). Retrieved May 6, 2024. https://www.nature.com/articles/d41586-023-04045-8.

4. Nature. Nuclear-fusion lab achieves 'ignition': what does it mean? (December 13, 2022) Retrieved May 6, 2024. https://www.nature.com/articles/d41586-022-04440-7

5. Chen, KY., Maiwald, J., Schauer, P.A. et al. Electrochemical loading enhances deuterium fusion rates in a metal target. Nature 644, 640–645 (2025). https://doi.org/10.1038/s41586-025-09042-7

6. Casey Daniel T., et. al., Towards the first plasma-electron screening experiment. Frontiers in Physics. Vol 10 (2022). doi: 10.3389/fphy.2022.1057603

7. Shaviv NJ, Shaviv G. Obtaining the electrostatic screening from first principles. Nucl Phys A (2003) 719:C43–C51. doi:10.1016/S0375-9474(03)00956-4

8. Däppen W, Mussack K. Dynamic screening in solar and stellar nuclear reactions. Contrib Plasma Phys (2012) 52:149–52. doi:10.1002/ctpp.201100099

9. Spitaleri C, Bertulani CA, Fortunato L, Vitturi A. The electron screening puzzle and nuclear clustering. Phys Lett B (2016) 755:275–8. doi:10.1016/j.physletb.2016.02.019

10. B. Steinetz, *et al*., NASA TP-20205001616, and Phys. Rev. C 101, 044610 (2020).

11. V. Pines, *et al*.., NASA TP-20205001617, and Phys. Rev. C 101, 044609 (2020).

12. Greife *et al*., Z. Phys. A 351 107 (1995).

13. Berlinguette, C.P. *et al*., "Revisiting the cold case of cold fusion." Nature 570, 45–51 (2019). https://doi.org/10.1038/s41586-019-1256-6.

14. Raiola, F. *et al*., (2006). Enhanced d(d,p)t fusion reaction in metals. In: Fülöp, Z., Gyürky, G., Somorjai, E. (eds) The 2nd International Conference on Nuclear Physics in Astrophysics. Springer, Berlin, Heidelberg. https://doi.org/10.1007/3-540-32843-2_11

15. Y. Iwamura , T. Itoh , J. Kasagi , S. Murakami , M. Saito. Excess Energy Generation using a Nano-sized Multilayer Metal Composite and Hydrogen Gas.

16. Y. Iwamura, T Itoh, J. Kasagi, A. Kitamura, A. Takahashi, K. Takahashi, R. Seto, T. Hatano, T. Hioki, T. Motohiro, M. Nakamura, M. Uchimura, H. Takahashi, S. Sumitomo,

Y. Furuyama, M. Kishida, H. Matsune. Anomalous Heat Effects Induced by Metal Nano-composites and Hydrogen Gas.

17. J. Kasagi. Low energy nuclear cross sections in metals Surf. Coat. Technol , volume 201 , issue 19-20 , p. 8574 - 8578 Posted: 2007.

18. Strieder F, Rolfs C, Spitaleri C, Corvisiero P. Electron-screening effects on fusion reactions. Naturwissenschaften (2001) 88:461–7. doi:10.1007/s001140100267.

19. Konrad Czerski Deuteron-deuteron nuclear reactions at extremely low energies. Phys. Rev. C 106, L011601 – Published 12 July 2022.

20. Kasagi, J., *et al*., Low Energy Nuclear Fusion Reactions in Solids. in 8th International Conference on ColdFusion. 2000. Lerici (La Spezia), Italy: Italian Physical Society, Bologna, Italy.

21. T. Schenkel, *et al*., "Investigation of Light Ion Fusion Reactions with Plasma Discharges," J. Appl. Phy., vol. 126, pp. 203302 (2019). doi: 10.1063/1.5109445

22. T.L. Benyo and L.P. Forsley, "MCNP Fusion Modeling of Electron-Screened Ions," 2021 MCNP® User Symposium, Los Alamos National Laboratory

23. G. Lehner and F. Pohl, "Reaktionsneutronen als Hilfsmittel der Plasmadiagnostik", Z. Physik 207 83-104 (1967).

24. W. A. Fisher, S. H. Chen, D. Gwinn, and R. R. Parker, Measurement of the D–D fusion neutron energy spectrum and variation of the peak width with plasma ion temperature. Phys. Rev. A 28, 3121 – Published 1 November 1983.

25. R.M. White and D.A. Resler, "ENDF/B-VI MT103, Nucl. Data Sheets, June 1991.

26. J. A. Kulesza, et al.  MCNP® Code Version 6.3.0 Theory & User Manual. Los Alamos National Laboratory Tech. Rep. LA-UR-22-30006, Rev. 1. Los Alamos, NM, USA. September 2022.

27. IEEE Spectrum (2022).  NASA's New Shortcut to Fusion Power. Retrieved July 7, 2025 from https://spectrum.ieee.org/lattice-confinement-fusion

28. Renaudin G., Yvon K., Dolukhanyan S.K., Aghajanyan N.N., Shekhtman V.S. "Crystal structures and thermal properties of titanium carbo-deuterides as prepared by combustion synthesis." J. Alloys Compd. 2003, 356/, 120-127.

29. H.L. Yakel, Jr., Thermocrystallography of Higher Hydrides of Titanium and Zirconium, Acta Cryst. 11, 46 (1958).

30. Neutron Fluence Measurements, International Atomic Energy Agency, Vienna, 1970. STI/DOC/10/107

31. B.E. Watt, Energy spectrum of neutrons from thermal fission of $^{235}$U, Phys. Rev. 87 (1952) 1037.

32. Anže Pungerčič, Dušan Čalič, Luka Snoj, Computational burnup analysis of the TRIGA Mark II research reactor fuel, Progress in Nuclear Energy, Volume 130, 2020, 103536, doi: 10.1016/j.pnucene.2020.103536.

33. J. Hall, "Mapping the Gamma Flux Level Around the UMR Reactor Core for Irradiation Applications" (2005). Opportunities for Undergraduate Research Experience Program (OURE). 238. https://scholarsmine.mst.edu/oure/238

34. M. Avila-Rodriguez *et al*., Neutron and gamma dose equivalent rates in the vicinity of a self-shielded PET cyclotron after an upgrade with increased beam current. Journal of Nuclear Medicine May 2011, 52 1429

35. M. Zmeškal *et al*., On radiation situation in vicinity of PET production cyclotron. EPJ Web of Conferences 308, 06002 (2024) doi.org/10.1051/epjconf/202430806002

36. Brad D. Jeffries, *et al*., Characterization of the neutron flux during production of $^{18}$F at a medical cyclotron and evaluation of the incidental neutron spectrum for neutron damage studies, Applied Radiation and Isotopes, Volume 154, 2019, 108892, ISSN 0969-8043, https://doi.org/10.1016/j.apradiso.2019.108892.

37. A. K. Gillespie, "Triton generation pathways: enhanced shielding from quantum nucleonics in metal hydrides." 51$^{st}$ Winter Colloquium on the Physics of Quantum Electronics. January 10-14, 2022. Snowbird, UT.

38. R. V. Duncan, "Quantum Nucleonics and Novel Nuclear Science." 51$^{st}$ Winter Colloquium on the Physics of Quantum Electronics. January 10-14, 2022. Snowbird, UT.

39. QuantaSmart for the Tri-Carb Liquid Scintillation Analyzer, Perkin Elmer QuantaSmart Reference Manual. Singapore, 2017.

40. State of Texas and Texas Tech University. (2022). Department of Physics and Astronomy. TTU. Retrieved July 18, 2024, from https://www.depts.ttu.edu/phas/cees/Instruction/PHYS_5300-19/Reading/Quantulus_GCT_6220_SOP.pdf

41. State of Texas and Texas Tech University. (November 7, 2022). Department of Physics and Astronomy. TTU. Retrieved November 9, 2022, from www.depts.ttu.edu/phas/cees/Ref/Q_Tritium.pdf

42. Ultima Gold uLLT scintillation cocktail, 6013681. Perkin Elmer Radiometric Reagents Guide 2010-2011, p. 138 https://per-form.hu/wp-content/uploads/2015/08/perkinelmer-Radiometric-reagents.pdf

43. R.P. Thorn, *et al*., "A quantitative light-isotope measurement system for climate and energy applications." Int J Mass Spectrom. 2021, 464.

44. State of Texas and Texas Tech University. (November 7, 2022). Department of Physics and Astronomy. TTU. Retrieved November 9, 2024, from https://www.depts.ttu.edu/phas/cees/

45.     Information Technology Division of Texas Tech University. (March 17, 2024). Research Data Management Services. Retrieved April 22, 2024, from https://www.depts.ttu.edu/library/digitalServices/index.php